\documentclass[RNAAS]{aastex62}
\usepackage{url}

\providecommand{\gaia}{Gaia}
\providecommand{\gdr}[1]{GDR{#1}}
\providecommand{\oum}{{1I/`Oumuamua}}
\providecommand{\tenc}{\ensuremath{t_{\rm enc}}}
\providecommand{\denc}{\ensuremath{d_{\rm enc}}}
\providecommand{\venc}{\ensuremath{v_{\rm enc}}}

\providecommand{\dencmed}{\ensuremath{d_{\rm enc}^{\rm med}}}
\providecommand{\vencmed}{\ensuremath{v_{\rm enc}^{\rm med}}}

\providecommand{\kms}{\ensuremath{\textrm{km/s}}}
\providecommand{\ra}{\ensuremath{\alpha}}
\providecommand{\dec}{\ensuremath{\delta}}
\providecommand{\dist}{\ensuremath{d_{\rm cur}}}
\usepackage[UKenglish]{isodate}

\begin{document}

\title{Future stellar flybys of the Voyager and Pioneer spacecraft}

\correspondingauthor{Coryn Bailer-Jones}
\email{calj@mpia.de}

\author{Coryn A.L.\ Bailer-Jones}
\affil{Max Planck Institute for Astronomy, K\"onigstuhl 17, 69117 Heidelberg, Germany}

\author[0000-0003-0774-884X]{Davide Farnocchia}
\affiliation{Jet Propulsion Laboratory, California Institute of Technology, 4800 Oak Grove Drive, Pasadena, CA 91109, USA}

\section{} 

\vspace*{-3em}
\begin{center} 
  {\em Extended version of our article published on 3 April 2019 in\\
    Research Notes of the American Astronomical Society, 3, 59}
\end{center}

\vspace*{1em}
\section*{Abstract}

The Pioneer 10 and 11 and Voyager 1 and 2 spacecraft, launched in the 1970s, are heading out of the solar system. Using the astrometric and radial velocity data from the second Gaia data release, we integrate the trajectories of 7.4 million stars, and the spacecraft, through a Galactic potential in order to identify those stars the spacecraft will pass closest to. The closest encounters for all spacecraft take place at separations between 0.2 and 0.5\,pc within the next million years. The closest encounter will be by Pioneer 10 with the K8 dwarf HIP 117795, at 0.23\,pc in 90\,kyr at a high relative velocity of 291\,\kms.

\vspace*{1em}
\section*{Introduction}

Following their encounters with the outer planets in the 1970s and 1980s, Pioneers 10 and 11 and Voyagers 1 and 2 are now on escape trajectories out of the solar system.  Although they will cease to operate long before encountering any stars (the Pioneers already have),
it is nonetheless interesting to ask which stars they will pass closest to in the next few million years. We answer this here using the accurate three-dimensional (3D) positions and 3D velocities of 7.2 million stars in the second \gaia\ data release (\gdr{2}, \citealt{2018A&A...616A...1G}), supplemented with radial velocities for 222\,000 additional stars obtained from Simbad\footnote{\url{http://simbad.u-strasbg.fr/}}.
A fifth spacecraft, New Horizons (which encountered Pluto in July 2015; \citealt{2015Sci...350.1815S}) has also achieved escape velocity to leave the solar system. However, as it may still undergo a manoeuvre to encounter another Edgeworth--Kuiper belt object (it already manoeuvred to intercept Arrokoth in January 2019), we have not included it in the current study.

\vspace*{1em}
\section*{Method}

To find encounters, we adopt the same method we used for tracing the possible origin (and future encounters) of the interstellar object \oum\ \citep{2018AJ....156..205B}.  We determine the asymptotic trajectories of the four spacecraft by starting from their ephemerides taken from JPL's Horizons system\footnote{\url{https://ssd.jpl.nasa.gov/horizons.cgi}}, which are at the of end 2030\,CE for Voyagers 1 and 2 and at the end of 2049\,CE for Pioneers 10 and 11.  We then propagate these numerically through the solar system to the year 2900\,CE, and finally extrapolate the Keplerian orbit to determine the asymptotic velocity vector for each spacecraft.

Using a linear motion approximation \citep{2015A&A...575A..35B} for the motion of these spacecraft
and the 7.4 million stars from \gdr{2}/Simbad, we identify those stars that will approach within 15\,pc of each spacecraft ($\sim 4500$ stars in each case).  We then integrate the orbits of these stars and the spacecraft through a Galactic potential and identify close encounters. We use the same model for the Galaxy as in \cite{2015A&A...575A..35B}.
Statistics of the encounter time, separation, and velocity are obtained by resampling the covariance of the stellar data and integrating the orbits of the resulting samples.
From the distribution of these samples we report the median as the point estimate, and the 5th and 95th percentiles as a measure of the uncertainty.
The uncertainties on the asymptotic spacecraft trajectories are negligible compared to those of the stars, and are therefore neglected.

\vspace*{1em}
\section*{Results}

\begin{figure}
\begin{center}
\includegraphics[width=1.0\textwidth, angle=0]{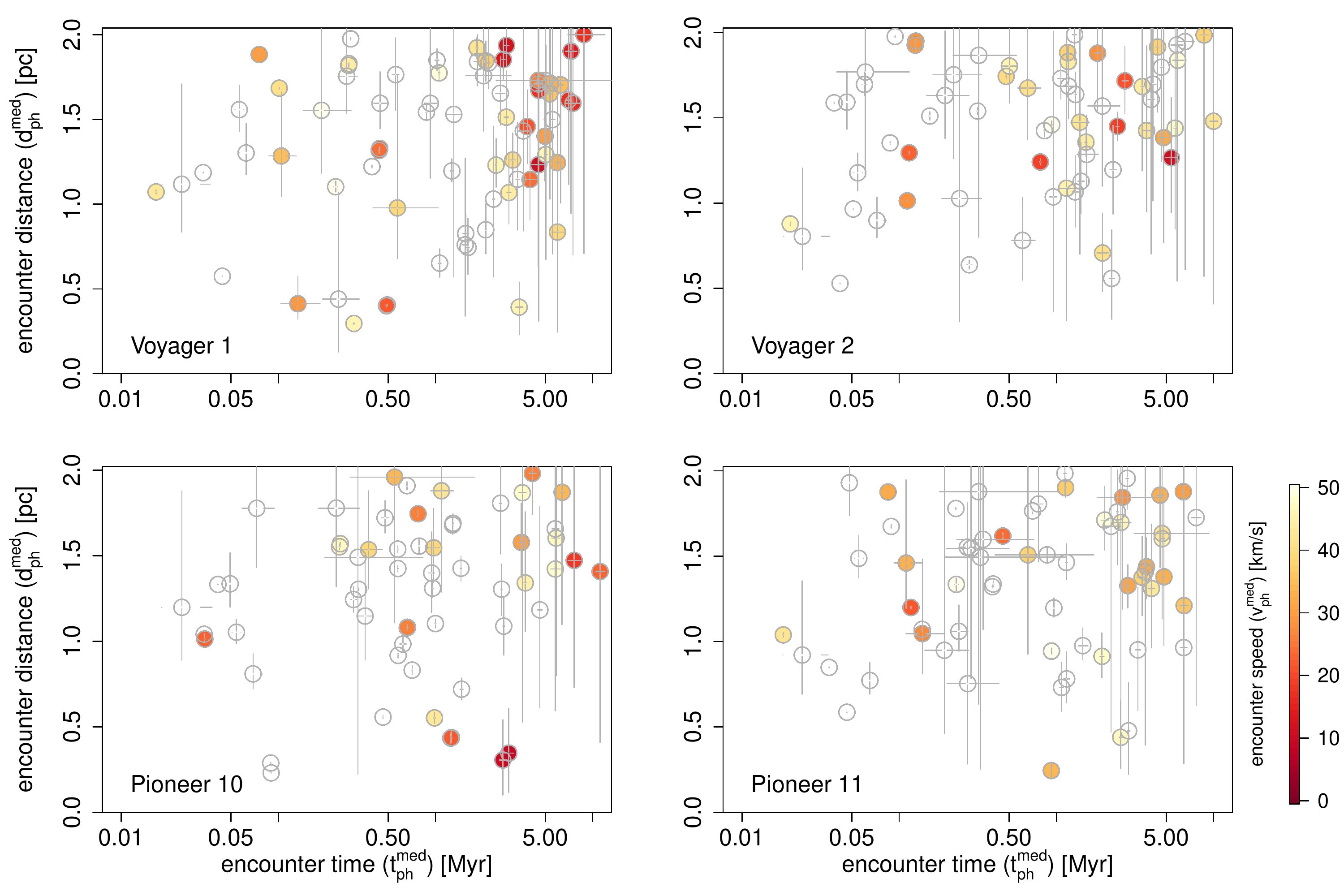}
\caption{Encounter parameters for the stars that approach within 2\,pc of each spacecraft. The point shows the median encounter time and distance (computed over the resampled stellar data); the error bars show the extent of the 5th and 95th percentiles (the error bars are smaller than the points in some cases). The colour indicates the relative velocity at the encounters. White circles have median encounter velocities greater than 50\,\kms. The horizontal axis is log time in the future to the encounter (the latest being for Pioneer 10 in 11.1\,Myr).
\label{fig:encounters}
}
\end{center}
\end{figure}

For each spacecraft, we identify about 60 stars that the spacecraft will approach within 2\,pc
(and about 10 stars within 1\,pc). The encounter parameters are plotted in Figure~\ref{fig:encounters}.
These and other data are provided in the accompanying tables.\footnote{\url{http://www.mpia.de/homes/calj/voyager_gdr2.html}} Some of these encounter are unreliable due to implausibly large radial velocities, possibly poor astrometric solutions, and/or undetected binarity. Very bright stars like Sirius and Alpha Centauri are not in \gdr{2} so are not covered by our study.

\begin{table*}
\begin{center}
\caption{Notable flybys. For each spacecraft the first and closest encounters are shown in the first and second lines respectively. ``Asymptote'' is the barycentric ICRF direction (degrees) and velocity (\kms) of the spacecraft as it leaves the solar system.  (For comparison, stars typically encounter the Sun at a relative velocity of 45\,\kms.)
 \tenc, \denc, \venc\ are the time from now, the star--spacecraft separation, and the relative velocity, respectively, of the encounters. ``med'' is the median of the distribution over the samples, with uncertainties represented by the 5th and 95th percentiles of the distributions.
\dist\ is the current distance to the star, taken from \cite{2018AJ....156...58B} (the precision is typically much higher than the rounding precision reported here).
SpT is the spectral type (from Simbad if available; those marked ``late K'' have been estimated from the absolute magnitude and colour).
The list of all encounters out to $\dencmed<2$\,pc is available from \url{http://www.mpia.de/homes/calj/voyager_gdr2.html}.
\label{tab:encounters}
}
\begin{tabular}{*{11}{r}}
\hline
Spacecraft \&      & Star                     & \multicolumn{3}{c|}{\tenc [kyr]} &  \multicolumn{3}{c}{\denc [pc]} &  \vencmed  &  \dist & SpT \\
\cline{3-5}\cline{6-8}
asymptote           &   name or \gdr{2} source ID      & med & 5\% & 95\% & med & 5\% & 95\% & [\kms] & [pc] & \\
\hline
{\bf Voyager 1}  &  Proxima Centauri & 16.7   &   16.5  &    16.9 &   1.072  &    1.061 &  1.083  &   43.2 &    1.3 &  M5.5V \\ 
$\ra=262.8760$ & TYC 3135-52-1 & 302.7  &   299.5  &   306.1 &   0.296 &   0.289 &    0.302 &   46.5 & 14.4 & M3V \\ 
$\dec=12.3199$ & 2091429484365218432 &  3405.3  &  3224.4  &  3595.7 &  0.392  &  0.229 &    0.540 &  45.9 & 159.5 & late KV \\ 
$v=16.6048$      & HD 28343 & 487.5   &    483.2    & 492.3   & 0.400 &    0.395 &  0.407 &    22.5 & 11.2 & M0.5V \\ 
                         & Gl 445 & 44.0   &   44.0 &     44.1 &   0.575  &  0.574 &   0.576 &  116.0 & 5.3 &  M4.0V   \\ 
\hline
{\bf Voyager 2}    & Proxima Centauri &   20.3 &    20.2  &    20.3 &   0.878 &    0.867 &    0.890 &    46.3 &    1.3 &   M5.5V \\ 
$\ra=316.2717$   & Ross 248      & 42.0 &     41.9  &    42.2 &   0.529 &    0.528 &   0.531  &    72.3  &  3.2 &  M5.0V \\ 
$\dec=-67.5491$ & 4370380741264455296 & 2245.6  &  2168.4 &   2334.7 &   0.558 &    0.319 &   0.843 &  69.9 & 160.5 & late KV \\ 
$v=14.8550$         & h$^1$ Sgr & 1965.4  &  1810.8 &   2141.0 &   0.708 &    0.480 &    0.941 &   40.9 & 82.2 & A1mA2-F0 \\ 
\hline
{\bf Pioneer 10}  & Ross 248    & 33.8     &     33.7  &    33.8    &  1.041  &  1.039 &   1.044 &   86.3 & 3.2 & M5.0V \\ 
  $\ra=83.4169$    &   HIP 117795 &  90.0  &    89.7  &    90.3 &   0.231 &    0.230 &    0.232 &   290.7 & 26.8  & K8V \\ 
$\dec=26.2171$ & HD 52456   & 1250.5 &   1238.4 &   1262.4 &   0.434 &    0.403 &    0.464 &    22.0 & 28.2 & K2V \\ 
$v=11.3149$       & Proxima Centauri &  34.1  &    33.6  &    34.4  &  1.013  &  0.990 &   1.033 &   23.4 & 1.3 &  M5.5V \\   
\hline
{\bf Pioneer 11}   & Proxima Centauri       & 18.3 &     18.1  &    18.5 &   1.040 &    1.027 &  1.052 &   41.8 &   1.3 &  M5.5V \\ 
$\ra=291.8277$   & TYC 992-192-1 &   928.3  &   920.0  &   937.4 &   0.245 &   0.200  &  0.292 &  33.6 & 31.9 & late KV \\ 
$\dec=-9.2212$ & 4544730574956793856 & 2568.1 &   2504.2 &   2640.7 &   0.439 &   0.257 &    0.653 &  47.2 & 123.9 & late KV\\ 
$v=10.4439$        & Gl 445 & 46.5  &    46.4  &    46.6 &   0.586 &    0.585  &    0.588&   109.9 &   5.3 &  M4.0V   \\ 
& $\delta$ Scuti      &  1163.4 &   1110.5 &   1216.1 &   0.782 &  0.641  &  0.942 &   51.3  &  61.0 &  F2II-III  \\ 
\hline
\end{tabular}
\end{center}
\end{table*}

The most interesting encounters that we find are listed in Table~\ref{tab:encounters} (this only shows
stars with reliable data). This table also gives the asymptotic velocity of the spacecraft relative to the solar system barycentre. For orientation: At a velocity of 10\,\kms, a spacecraft will travel 1\,pc every 100\,kyr.

The flybys of Voyager 1 with Gl 445  (AC +79 3888) and Voyager 2 with Ross 248 have been previously  identified (e.g.\ on the JPL website\footnote{\url{https://voyager.jpl.nasa.gov}}) but are now characterized more precisely in the present study using \gdr{2}.
Proxima Centauri (Gaia DR2 5853498713160606720) is the first flyby (within about 1\,pc) for three of the spacecraft, because that star is the current nearest neighbour to the Sun (at 1.3\,pc).
All four spacecraft will encounter it at a distance of about 1\,pc, so these are not close encounters.
Note that we have not taken into account Proxima Centauri's acceleration due to 
Alpha Centauri AB in our simulations. We discuss now some of these encounters.

\paragraph{{\bf Voyager 1}}
Voyager 1's closest flyby, TYC 3135-52-1 (Gaia DR2 2051984436005124480), lies above the main sequence in the colour-magnitude diagram, suggesting it may be a binary, in which case its astrometry may be unreliable.
Gaia DR2 2091429484365218432 and HD 28343 (Gaia DR2 145421309108301184) are the second and third closest encounters respectively. 
Gl 445 (Gaia DR2 1129149723913123456) is the earliest encounter within 1\,pc. 

\paragraph{{\bf Voyager 2}}
Ross 248 (Gaia DR2 1926461164913660160) is the closest encounter for Voyager 2. The second closest
is with Gaia DR2 4370380741264455296, even though that star is currently 160\,pc away.
Its fourth closest encounter will be with h$^1$ Sgr (Gaia DR2 6767920580693895808), an intrinsically bright and blue delta Scuti-type star.

\paragraph{{\bf Pioneer 10}}
HIP 117795 (Gaia DR2 2011565220332867584) is the closest encounter to Pioneer 10. Although its \gdr{2} radial velocity of $-285.87 \pm 0.41$\,\kms\ is suspiciously large (and based on just two observations),
especially when considering its modest tangential velocity of just 11\,\kms,
\cite{2016A&A...596A.116S}
report an independent and consistent radial velocity of  $-285.9 \pm 0.2$\,\kms, which is based on six measurements make over the course of four years.
\cite{2018A&A...616A..37B} found that this star will pass 1.05\,pc from the Sun 100 to 200 years later.
HD 152311 (Gaia DR2 4127420626087054976) is a binary, 
so its astrometry is possibly incorrect.
This leaves HD 52456 (Gaia DR2 3153772873178431744) as the next closest encounter.
Ross 248 is the earliest encounter, 
and Pioneer 10 has its closest approach to Proxima Centauri at essentially the same time (the median time is just 300 years later).

\paragraph{{\bf Pioneer 11}}
Pioneer's 11 closest flyby is with TYC 992-192-1 (Gaia DR2 4490721567368465408).
Its next closest is with Gaia DR2 454473057495679385.
Gl 445 appears again here, as the fourth closest encounter.
It is interesting that in 1.2\,Myr, Pioneer 11 will pass 0.8\,pc from the star $\delta$ Scuti (Gaia DR2 4155413814250528128), the prototype of the eponymous variability class. 
Although this star has an inflated unit weight error (5.4) in the \gdr{2} astrometry, its parallax and proper motion are similar to the Hipparcos values, so its astrometry is probably reasonably reliable.

\vspace*{1em}
\section*{Conclusions}

We have identified stars from \gdr{2} that the Pioneer 10 and 11 and Voyager 1 spacecraft will pass within 0.2--0.3\,pc in the next million years. Voyager 2's closest approach is much sooner (40\,kyr), but at a larger separation (0.5\,pc).  Statistically, a spacecraft will encounter stars within a given distance at approximately the same rate as the Sun does, which \cite{2018A&A...616A..37B} inferred to be one star within 1\,pc every 50\,kyr.  This inferred rate is an extrapolation from the \gdr{2} data and so predicts many more stars than found here, because most stars do not have the necessary six-dimensional phase space data in \gdr{2}.  This rate scales quadratically with encounter distance (i.e.\ one star within 0.1\,pc every 5\,Myr). As the spacecraft are not leaving the Galaxy, it is inevitable that the spacecraft will pass much closer to some stars on longer timescales than found here. These stars cannot yet be identified because of the limiting magnitude (and thus distance horizon) of the available data. Future data releases from \gaia\ and other surveys that provide radial velocities for more -- especially fainter -- stars could reveal specific, closer flybys.
The timescale for the collision of a spacecraft with a star is of order $10^{20}$ years, so the spacecraft have a long future ahead of them.

\acknowledgments

This study has used data from the European Space Agency (ESA) mission
\gaia\ (\url{http://www.cosmos.esa.int/gaia}), processed by the \gaia\
Data Processing and Analysis Consortium. 
Funding for the DPAC has been provided by national institutions, in particular the institutions participating in the \gaia\ Multilateral Agreement.
D.\ Farnocchia conducted this research at the Jet Propulsion Laboratory, California Institute of Technology, under a contract with NASA.

\bibliography{voyager_gdr2}

\end{document}